\documentclass{article}
\usepackage{arxiv}
\usepackage[numbers,sort&compress]{natbib}
\usepackage[utf8]{inputenc} 
\usepackage[T1]{fontenc}    
\usepackage{ dsfont }
\usepackage{hyperref}       
\usepackage{url}
\usepackage{textcomp}
\usepackage{ gensymb }
\usepackage[ruled,vlined]{algorithm2e}
\usepackage{algorithmic}
\usepackage{booktabs}       
\usepackage{amsfonts}       
\usepackage{nicefrac}       
\usepackage{microtype}      
\usepackage{lipsum}
\usepackage{float}
\usepackage{graphicx}
\usepackage{multirow}
\usepackage{array}
\usepackage{comment}
\usepackage{longtable}
\usepackage{ragged2e}
\usepackage{subfig}
\usepackage{amsmath}
\title{Transfer learning and Local interpretable model agnostic based visual approach in Monkeypox Disease Detection and Classification: A Deep Learning insights}

\author{
  Md Manjurul Ahsan \\
  Department of Industrial and Systems Engineering\\
  University of Oklahoma\\
  Norman, Oklahoma-73071 \\
  \texttt{ahsan@ou.edu} \\
   \And   
  Tareque Abu Abdullah \\
  Department of Computer Science\\
  Lamar University\\
  Beaumont, TX-77710 \\
  \texttt{tabdullah@lamar.edu} \\
   \And
 Md Shahin Ali \\
  Department of Biomedical Engineering\\
  Islamic University\\
  Kushtia-7003, Bangladesh\\
  \texttt{shahin@std.iu.ac.bd}\\
  \And
    Fatematuj Jahora \\
  Child Development and Social Relationship\\
  Government College of Applied Human Science\\
  Azimpur, Dhaka-1205, Bangladesh \\
  \texttt{fatema.cdsr@gmail.com} \\
  \And
    Md Khairul Islam \\
  Department of Biomedical Engineering\\
  Islamic University\\
  Kushtia-7003, Bangladesh \\
  \texttt{khairul@bme.iu.ac.bd} \\
   \And
    Amin G. Alhashim \\
  Department of Industrial and systems Engineering\\
  Arizona State University\\
  7271 E. Sonoran Arroyo Mall Mesa, AZ-85212 \\
  \texttt{alhashim.asu@gmail.com} \\
   \And
Kishor Datta Gupta \\
  Department of Computer and Information Science\\
  Clark Atlanta University\\
  Atlanta, GA-30314, USA\\
  \texttt{kgupta@cau.edu}} 

\begin{document}

\maketitle
\begin{abstract}
The recent development of Monkeypox disease among various nations poses a global pandemic threat when the world is still fighting Coronavirus Disease-2019 (COVID-19). At its dawn, the slow and steady transmission of Monkeypox disease among individuals needs to be addressed seriously. Over the years, Deep learning (DL) based disease prediction has demonstrated true potential by providing early, cheap, and affordable diagnosis facilities. Considering this opportunity, we have conducted two studies where we modified and tested six distinct deep learning models—VGG16, InceptionResNetV2, ResNet50, ResNet101, MobileNetV2, and VGG19--using transfer learning approaches. Our preliminary computational results show that the proposed modified InceptionResNetV2 and MobileNetV2 models perform best by achieving an accuracy ranging from 93\% to 99\%. Our findings are reinforced by recent academic work that demonstrates improved performance in constructing multiple disease diagnosis models using transfer learning approaches. Lastly, we further explain our model prediction using Local Interpretable Model-Agnostic Explanations (LIME), which plays an essential role in identifying important features that characterize the onset of Monkeypox disease.

\end{abstract}
\section{Introduction}\label{sec1}

Monkeypox, also known as Monkeypox Virus (MPXV) disease, is caused by infection with the virus of the same name and is usually found in monkeys~\cite{patrono2020monkeypox}. Monkeypox usually occurs in Africa, Central and West Africa, and Asia~\cite{parker2007human}. Although it can infect any mammal, the virus spreads to humans primarily through the bite of an infected animal like a bat or primate such as a monkey~\cite{wong2007bats}. There are several early monkeypox symptoms, including muscle pain, headache, fatigue, and fever. The disease almost resembled chickenpox, smallpox, and measles. It can be distinguished from the appearance of swollen glands, usually present behind the ear, below the jaw, neck, or groin, before the onset of the rash~\cite{cdcsymptoms}. Even though the virus is not life-threatening, it causes complications in severe cases, including sepsis, pneumonia, and loss of eye vision~\cite{reynolds2017improving}. 

Although monkeypox does not affect humans very often, the slight chance of affecting it should still make people think twice about their exposure to monkeys and rodents — especially in certain parts of the world where outbreaks of this disease happen more often. According to the Centers for Disease Control and Prevention (CDC), a new strain of monkeypox disease will infect millions globally in the coming years~\cite{fauci2005emerging}.

Monkeypox was initially found in 1958 and was rediscovered in the Republic of Congo in West Africa in 2014~\cite{africacase50}. While it may not be as well-known as the likes of Ebola or Zika, the monkeypox virus has the potential to become just as large of a global health threat as those two if enough is not done to stop it from spreading further.
Recently, the virus has been slowly spreading, and the number of Monkeypox patients is increasing day by day. According to CDC, as of June 06, 2022, the virus had spread among 29 countries, and the number of Confirmed cases was around 1029~\cite{cdcstat}. Currently, there are no appropriate treatments available for Monkeypox disease~\cite{ahsan2022image,ahsan2022monkeypox}. However, two oral drugs, Brincidofovir and Tecovirimat, are recommended for urgent treatment, mainly used to treat the smallpox virus. Vaccination is the ultimate solution to prevent this disease~\cite{notreatment2022}.

The symptoms of Monkeypox, smallpox, and measles are almost similar making it difficult to identify without a proper laboratory test~\cite{heymann1998re}. The definitive way to diagnose the virus is to use electronic microscopy and test the skin lesions. In addition, the virus can be identified using Polymerase chain reaction (PCR) testing, which is heavily used to test for COVID-19~\cite{Monkeypoxagri2022}.
 PCR is usually used for laboratory tests. However, during the onset of COVID-19, we have seen that the PCR test kit cannot correctly diagnose COVID-19 patients around 40\% of the time~\cite{ahsan2021detecting} which means that multiple tests are required to increase accuracy. As an effect, if Monkeypox becomes another global pandemic, then it will be hard to facilitate enough tool kits to test both Monkeypox and COVID-19 patients. Additionally, many developed nations cannot afford such an expensive tool kit to use regularly~\cite{ahsan2020deep}.
 
In recent past Machine learning (ML) demonstrated promising results in medical imaging and disease diagnosis~\cite{ahsan2022machine} which means that diseases such as cancer, pneumonia, and COVID-19 can be detected without doctor intervention~\cite{ahsan2022machinedia}. As Monkeypox symptoms significantly affect the human skin, images containing infected human skin can be used to develop ML-based diagnosis model~\cite{ahsan2022image}.

Monkeypox disease detection using any ML or deep learning (DL) approach is not available at the time of writing. However, over the year, several studies  demonstrated that DL-based models can be a reliable option to detect Chickenpox and Measles disease, which have almost similar symptoms as Monkeypox~\cite{chae2018predicting,arias2020varicella,bhadula2019machine}. For example, Chae et al. (2018) employed a very deep neural network (DNN) and long-short term memory (LSTM) model to identify chickenpox, which outperforms the classic autoregressive integrated moving average (ARIMA) model~\cite{chae2018predicting}. Arias et al. (2020) developed a deep learning model to detect varicella-zoster, a virus that can cause damage to the eyes. Their proposed model can identify the virus up to  97\% accuracy. Bhadula et al. (2019) employed convolutional neural networks (CNN) to detect skin disorders. The authors use the CNN model to detect acne and lichen planus and achieved 96\% and 92\% accuracy, respectively~\cite{bhadula2019machine}. Sriwong et al. (2019) utilized the CNN approach to detect skin diseases and achieved 79.2\% accuracy. Skin diseases such as actinic keratoses, basal cells, and benign keratosis are some of the skin diseases that the authors wanted to detect throughout the study~\cite{sriwong2019dermatological}.

Based on the referenced literature, it can be infer that, DL techniques have been used extensively to detect various skin diseases caused mainly by various infections. Therefore, traditional CNN is a good approach for developing deep learning-based models to diagnose the Monkeypox disease diagnosis. In this work, we proposed and tested six distinct improved deep CNN models by adopting models, namely, VGG16~\cite{simonyan2014very}, InceptionResNetV2~\cite{szegedy2017inception}, ResNet50~\cite{akiba2017extremely,islam2022improving,islam2022aimproving}, ResNet101~\cite{he2016deep}, MobileNetV2~\cite{sandler2018mobilenetv2}, and VGG19~\cite{simonyan2014very}, using transfer learning approaches~\cite{jones2022applying}. The following is a summary of our contribution:
\begin{enumerate}
    \item Improvement and assessment of six distinct deep learning models (VGG16, InceptionResNetV2, ResNet50, ResNet101, MobileNetV2, and VGG19) for detecting Monkeypox disease using image data.
    \item Verify and interpret the model's performance employing Local Interpretable Model Agnostic Explanations (LIME).

\end{enumerate}

\section{Methodology}\label{methods}
This study aims to detect Monkeypox disease by developing CNN model using transfer learning approaches. In this work, we have used pre-trained deep learning architectures to extract essential features that are practically difficult to initially identify by visual inspection due to their similarities with other infectious diseases such as chickenpox and measles. We then fed our data through several layers, where the top most dense layer is used to detect the Monkeypox disease.
\subsection{Dataset}
\begin{figure}
    \centering
    \includegraphics[width=.75\textwidth]{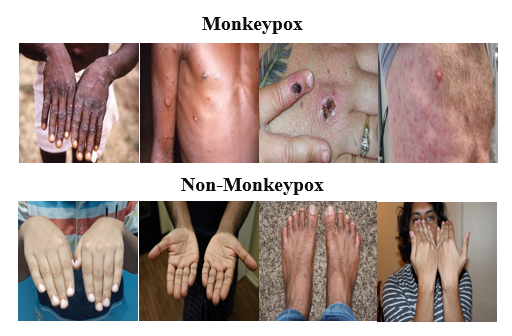}
    \caption{Images retrieved from open source data repositories to use in our proposed research.}
    \label{fig:si}
\end{figure}

The traditional approach of deep learning demands a significant amount of input data in order to train. The recent advancement of transfer learning approaches, on the other hand, demonstrates that a limited dataset could be used to train and develop a robust CNN model that can efficiently perform during the predictions~\cite{ahsan2021detecting}. At the time of writing, only one publicly available dataset exists, which is obtained from the \href{https://www.kaggle.com/datasets/dipuiucse/monkeypoxskinimagedataset}{Kaggle repository}. The assignment of the dataset that has been used during this study is presented in Table~\ref{tab:mpoxdata}. Our assessment of the data size differs for two different studies. For Study One, we have used 76 samples (43 Monkeypox and 33 Non-Monkeypox images), whereas, for Study Two, 818 samples are used, which contains 587 Monkeypox and 231 Normal images.
\begin{table}[ht]
 \caption{Monkeypox datasets used in this study.}
    \centering
    \begin{tabular}{ccccc}\toprule
         Dataset&	Label&	Train set&	Test set& Total\\\midrule
\multirow{2}{*}{Study One}	& Monkeypox&	34&	9& \multirow{2}{*}{76}\\
	&Normal&	26&	7\\\cmidrule{2-5}
	\multirow{2}{*}{Study Two}& Monkeypox&	469&	118&\multirow{2}{*}{818}\\
	&Normal&	185&	46\\\bottomrule

    \end{tabular}
   
    \label{tab:mpoxdata}
\end{table}
Figure~\ref{fig:si} presents a set of representative images of both Monkeypox and Non-Monkeypox individuals from the aforementioned datasets.

\subsection{Data Augmentation}

Data augmentation is a technique for increasing the size of a training dataset without acquiring new data from existing training data~\cite{ali2021enhanced}. Many computer vision applications have benefited from Deep CNN, which is sensitive to large amounts of data to reduce overfitting. An overfitting network detects a function with a significant variance to estimate the training dataset accurately. Affine transformations, deformations, or geometric distortions are often used to enrich data and enlarge sample sizes for deep neural networks~\cite{kwasigroch2017deep}. We have applied data augmentation to reduce the danger of overfitting, which involves random image alterations such as flipping, rotation height and weight shift, shearing, and zooming.

\noindent
\textbf{Flipping:} Flipping is performed on each image inside the training set. Vertical flips capture a distinctive property of medical images: reflection invariance in the vertical direction. Vertical inversions may not always correlate to images that are input. For input images, only horizontal flips of the source images are used. In contrast, a vertical mass flip would still result in a considerable weight.

\noindent
\textbf{Scaling:} Each I is scaled in either the x or y-axis; more precisely, affine transformation is used.
\(A\) = $\begin{pmatrix}
  s_x & 0\\ 
  0 & s_y
\end{pmatrix}$

\noindent
\textbf{Rotations:} Rotations are performed using the affine transformation as well, \(A\) = $\begin{pmatrix}
  cos\theta & -sin\theta\\ 
  sin\theta & cos\theta
\end{pmatrix}$ , where \(\theta\) is a value between \(0\) and \(360\) degrees.

\noindent
\textbf{Shearing:} Shear angle in degrees in a counter-clockwise direction scaled by 0.2 which is represented by the following affine transformation,
\(A\) = $\begin{pmatrix}
  1 & s \\
  0 & 1
\end{pmatrix}$. \(s\) defines the amount that I is sheared, and it is in the range of [0,  0.2].\\

\noindent
\textbf{Translation:} Simply moving the image on the X or Y axis is required for translation (or both). For example, it might be assumed that the image seems to have a black background outside its boundaries and translate the image accordingly. This enhancement can discover the vast majority of things anywhere inside the image. Thus, CNN is compelled to examine everywhere.

\subsection{Transfer learning approaches}
Recent research demonstrates the outstanding performance of deep CNN for the classification of image data with a high accuracy rate~\cite{ahsan2021detecting}. 
The development of deep CNN-based models that have established themselves as a promising tool for image classification and computer vision is gaining popularity day by day~\cite{ahsan2020face,ahsan2021evaluating}. Traditional CNN requires large amounts of data, whereas transfer learning can be applied to develop a model that can be adopted for other tasks~\cite{ahsan2021detecting}. During this study, the internal structure of the existing models was kept same to extract the important features from the Monkeypox images without further increasing any CNN-related complexity. Then the top most layers are modified to perform the classification. Throughout the course of this research, we modified six distinct deep-learning models, namely, VGG16, InceptionResNetV2, ResNet50, ResNet101, MobileNetV2, and VGG19.

\begin{figure}
    \centering
    \includegraphics[scale= .5]{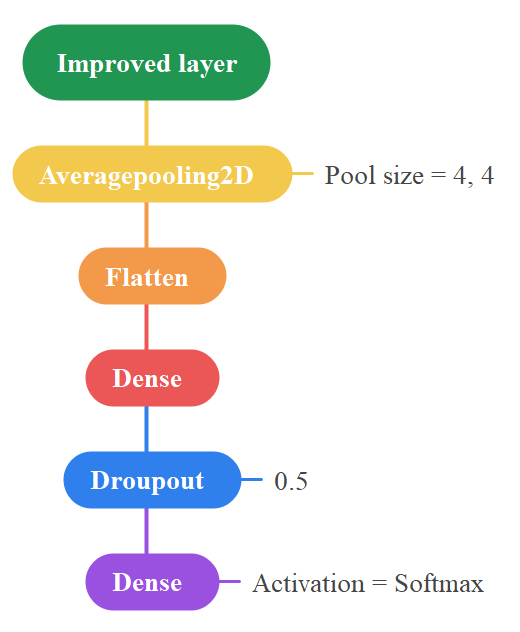}
    \caption{Modified layer}
    \label{fig:ml}
\end{figure}

\begin{figure*}
    \centering
    \includegraphics[width=\textwidth]{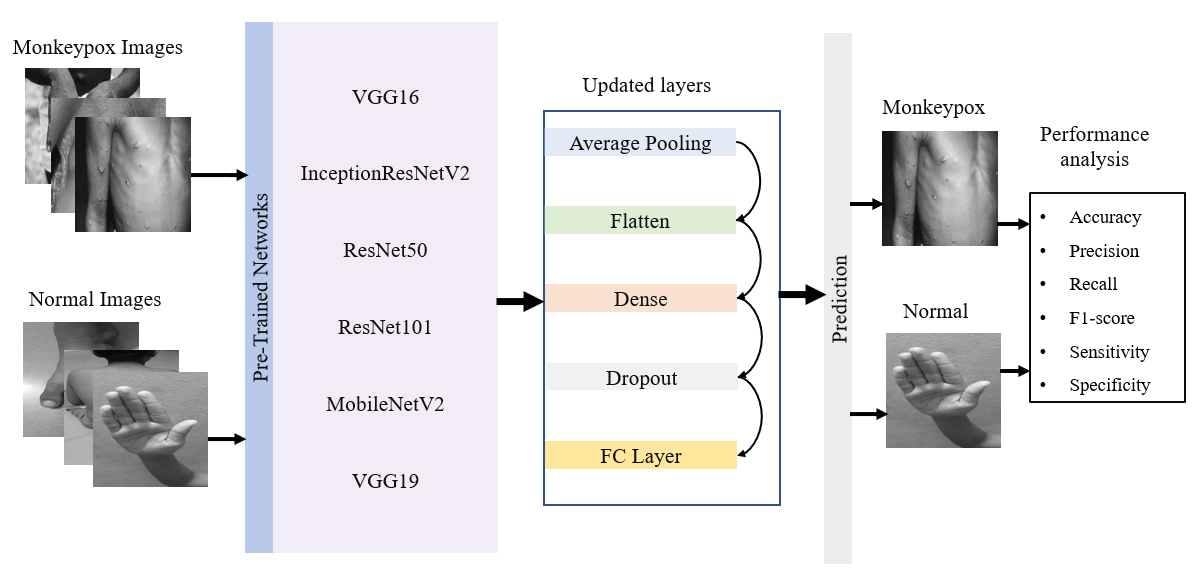}
    \caption{Flow diagram of the proposed models.}
    \label{fig:fd}
\end{figure*}

Figure~\ref{fig:fd} displays the flow diagram of the proposed models used during this study.

\subsection{Proposed architecture}
\begin{figure*}[ht]
    \centering
    \includegraphics[width=\textwidth]{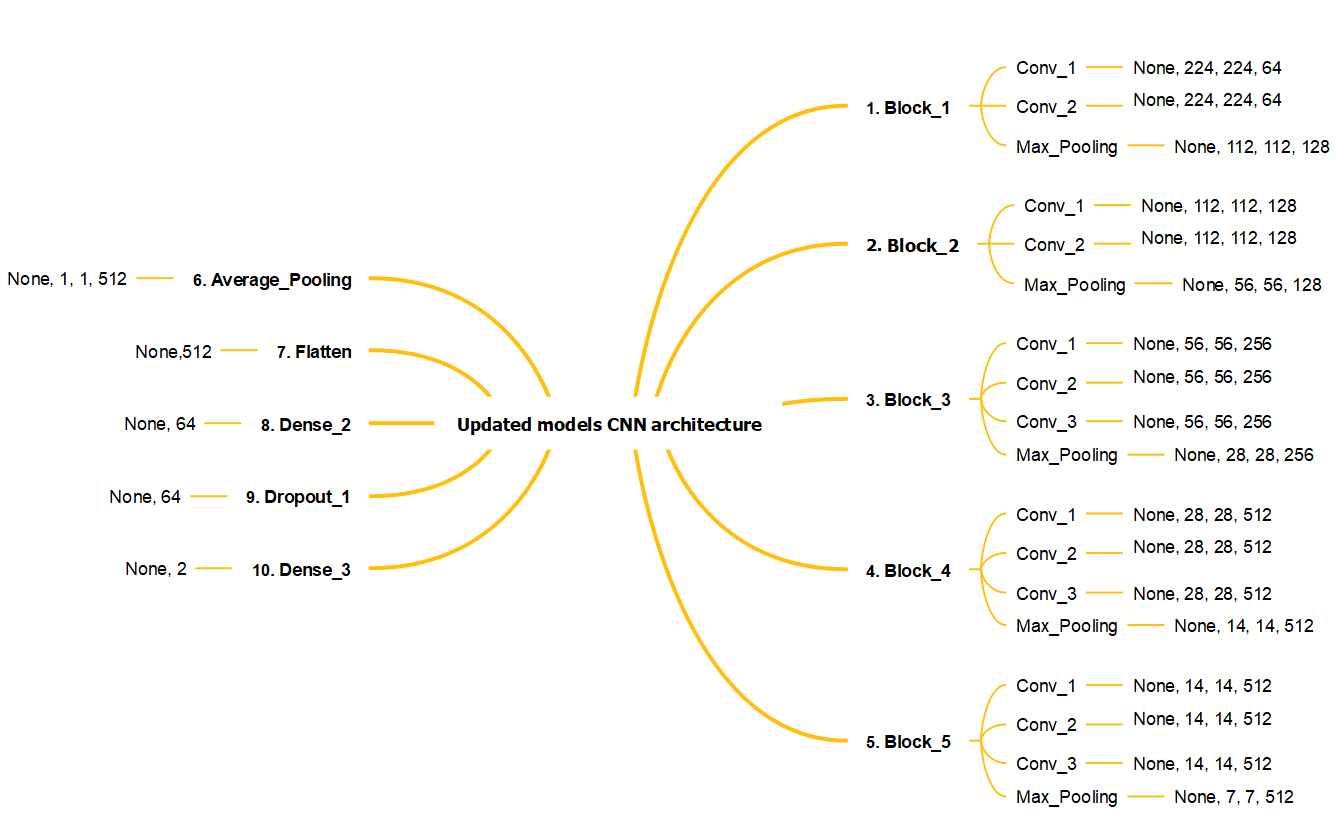}
    \caption{Updated VGG16 network implemented during this experiment.}
    \label{fig:vgg}
\end{figure*}

A pre-trained network is one that has been trained on a large dataset and can be applied to other datasets with comparable properties. A prime example is VGG16 networks, which have been adopted by many recent literatures to detect COVID-19 during the onset of the COVID-19 pandemic~\cite{ahsan2021detecting,narin2021automatic}. Figure~\ref{fig:vgg} shows a sample modified architecture that has been proposed during this study for VGG16.

The modified architecture generally follows the following three steps:
\begin{enumerate}
    \item The model is initially initialized with a pre-trained network without FC layers.
    \item Then a new FC layer is added with "Softmax" activation function and added to the top of the actual VGG16 model.
    \item Finally, during the training phase, the weight of the convolution is frozen so that the FC layer can be developed during the experiment.
\end{enumerate}
Similar procedure is followed to develop every transfer learning model used during this study. In this study, the primary modification of the model for all CNN architecture is developed as shown in Figure~\ref{fig:ml}.

Our proposed study optimized three parameters during the training phase: batch size, epochs, and learning rate. Manually tunning that parameter is time-consuming; therefore, we have used the grid search method to find the best results. Table~\ref{tab:ps} summarizes the initial and optimal parameters found during the experiment.

\begin{table*}[ht]
\caption{Parameter settings used during this study.}
    \centering
    \resizebox{\textwidth}{!}{
    \begin{tabular}{llllll}\toprule
      \multirow{2}{*}{Parameters}&	\multicolumn{2}{c}{Study One}& \multicolumn{2}{c}{Study Two}\\\cmidrule{2-5}
	&Initial parameter&	Optimize parameters&	Initial parameters&	Optimize parameters\\\midrule
Batch size&	5, 10, 20, 40&	10&	10, 20, 30, 40&	30\\
Epochs&	20, 30, 40, 50&	50&	30, 40, 50, 70&	50\\
Learning rate&	.001, .01, 0.1&	.001&	.001, .01, 0.1&	.001\\\bottomrule

    \end{tabular}}
    
    \label{tab:ps}
\end{table*}
\subsection{Experiment setup}

Adaptive momentum, also known as Adam, is employed to optimize the performance of our proposed model due to the fact that it has a reliable performance with regard to binary image classification~\cite{hameed2016back}.
The experiment was carried out using an office-grade laptop with a configuration of Windows 10, Core I7, and with 16 GB RAM facilities. The dataset is split into the following ratios, which is often common in machine learning domains—train set: test set= 80:20. The overall experimentation is run five times, and the final result is presented by averaging all five results. To evaluate the model's performance following statistical measurement is used~\cite{ahsan2020face}:
\begin{equation}\label{eq1}
    Accuracy = \frac{T_{p}+ T_{n}}{T_{p}+T_{n}+F_{p}+F_{n}}
\end{equation}
\begin{equation}\label{eq2}
    Precision = \frac{T_p}{T_p+F_p}
\end{equation}
\begin{equation}\label{eq3}
   Recall = \frac{T_p}{T_n+F_p}
\end{equation}
\begin{equation}\label{eq4}
 F1- score = 2\times\frac{\textrm{Precision}\times\textrm{ Recall}}{\textrm{Precision~+~Recall}}
\end{equation}
\begin{equation}\label{eq5}
    Sensitivity = \frac{T_p}{T_p+F_n}
\end{equation}
\begin{equation}\label{eq6}
    Specificity = \frac{T_n}{T_n+F_p}
\end{equation}
Here,
$T_p$ (True Positive) = Monkeypox infected individual identified as Monkeypox

 $T_n$ (True Negative) = Monkeypox infected individual identified as non-Monkeypox

 $F_p$ (False Positive) = Non-Monkeypox individual identified as Monkeypox

$F_n$ (False Negative) = Monkeypox individual identified as non-Monkeypox.

\section{Results}~\label{result}
\vspace{-6mm}
\subsection{Study One}
Equations~\ref{eq1}, \ref{eq2}, \ref{eq3}, \ref{eq4}, \ref{eq5} and \ref{eq6} were used in Study One to investigate each and every CNN models performance on both the train and test sets. In this particular scenario, InceptionResNetV2 and MobileNetV2 performed far better than any other models in terms of accuracy, precision, recall, F-1 score, sensitivity, and specificity (refer to Table~\ref{tab:std1}). On the other hand, ResNet50 showed the worst performance out of all the measurements.
\begin{table}[ht]
\caption{Monkeypox prediction performance of all models on the training set using various statistical measurements for Study One.}
    \centering
    \resizebox{\textwidth}{!}{
    \begin{tabular}{ccccccc}\toprule
         \multirow{2}{*}{Model}&	\multicolumn{5}{c}{Performance}\\\cmidrule{3-7}
	&Accuracy&	Precision&	Recall&	F1-Score&	Sensitivity&	Specificity\\\midrule
VGG16&	0.98&	0.98&	0.98&	0.98&	1.0&	0.96\\
InceptionResNetV2&
	1&	1&	1&	1&	1&	1\\
ResNet50&	0.56&	0.32&	0.57&	0.41&	1&	0\\
ResNet101&	0.68&	0.80&	0.68&	0.63&	1&	0.26\\
MobileNetV2&	1&	1&	1&	1&	1&	1\\
VGG19&	0.98&	0.98&	0.98&	0.98&	1&	0.96\\\bottomrule
 
    \end{tabular}}
    
    \label{tab:std1}
\end{table}
The results of the performance of all of the models with regard to the test set are presented in Table~\ref{tab:std12}. While the models ResNet50 and ResNet101 showed the least performance, InceptionResNetV2 and VGG19 demonstrated the best performance across all criteria.
\begin{table}[ht]
\caption{Monkeypox prediction performance of all models on the testing set using various statistical measurements for Study One.}
    \centering
    \resizebox{\textwidth}{!}{
    \begin{tabular}{ccccccc}\toprule
        \multirow{2}{*}{Model}&\multicolumn{5}{c}{Performance}\\\cmidrule{3-7}		
	&Accuracy&	Precision&	Recall&	F1-Score&	Sensitivity&	Specificity\\\midrule
VGG16&	0.81&	0.86&	0.81&	0.80&	1&	0.42\\
InceptionResNetV2&
		0.90&	0.88&	0.88&	0.77&	1& 0.95\\
ResNet50&	0.56&	0.32&	0.56&	0.40&	1&	0\\
ResNet101&	0.56&	0.32&	0.56&	0.40&	1&	0\\
MobileNetV2&	0.81&	0.82&	0.81&	0.81&	0.77&	0.85\\
VGG19&	0.93&	0.94&	0.94&	0.94&	1&	0.85\\\bottomrule

    \end{tabular}}
    \label{tab:std12}
\end{table}
\subsection{Confusion matrix}
\begin{figure*}[ht]
    \centering
    \includegraphics[width=\textwidth]{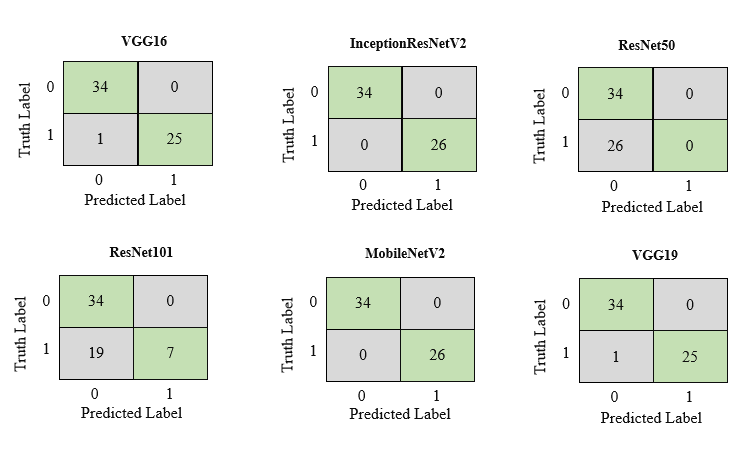}
    \caption{Study one confusion matrices for the train set.}
    \label{fig:con1}
\end{figure*}

The total prediction of the model can be represented using confusion matrices. Figure~\ref{fig:con1} illustrates the confusion matrices acquired for the train set during Study One. Both the InceptionResNetV2 and MobileNetV2 models accurately categorized all patients, as shown in the Figure~\ref{fig:con1}. Contrary, VGG16, ResNet50, ResNet101 and VGG19 misclassified 1, 26, 19 and 1 patients respectively. As a result, we can draw the conclusion that the performance of the ResNet50 model is the worst among all of the models based on the assessment of the confusion matrices.

\begin{figure*}[ht]
    \centering
    \includegraphics[width=\textwidth]{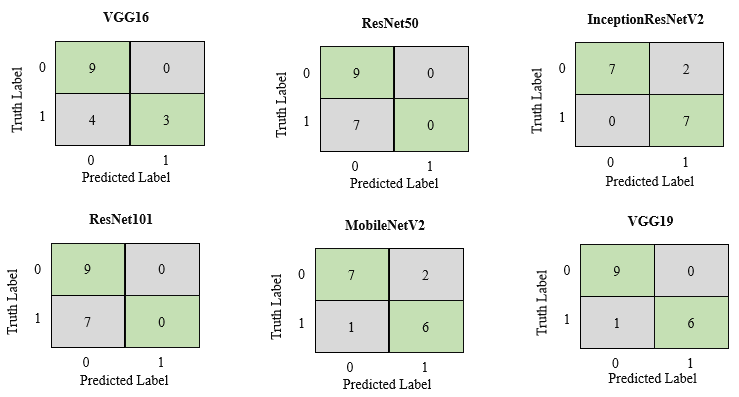}
    \caption{Study one confusion matrices for the test set.}
    \label{fig:con12}
\end{figure*}

Figure~\ref{fig:con12} displays the confusion matrices for Study One on the test set. Models InceptionResNetV2 displayed the best performance by misclassifying only two patients, while both ResNet50 and ResNet101 demonstrated the worst performance by misclassifying seven patients. Considering the performance on both the train and test set, it can be inferred that the performance of modified InceptionResNetV2 remains constant for both train and test set, ultimately demonstrated best performance compared to all other models.
\subsection{Models’ performance during training}
\begin{figure*}[ht]
    \centering
    \includegraphics[width=\textwidth]{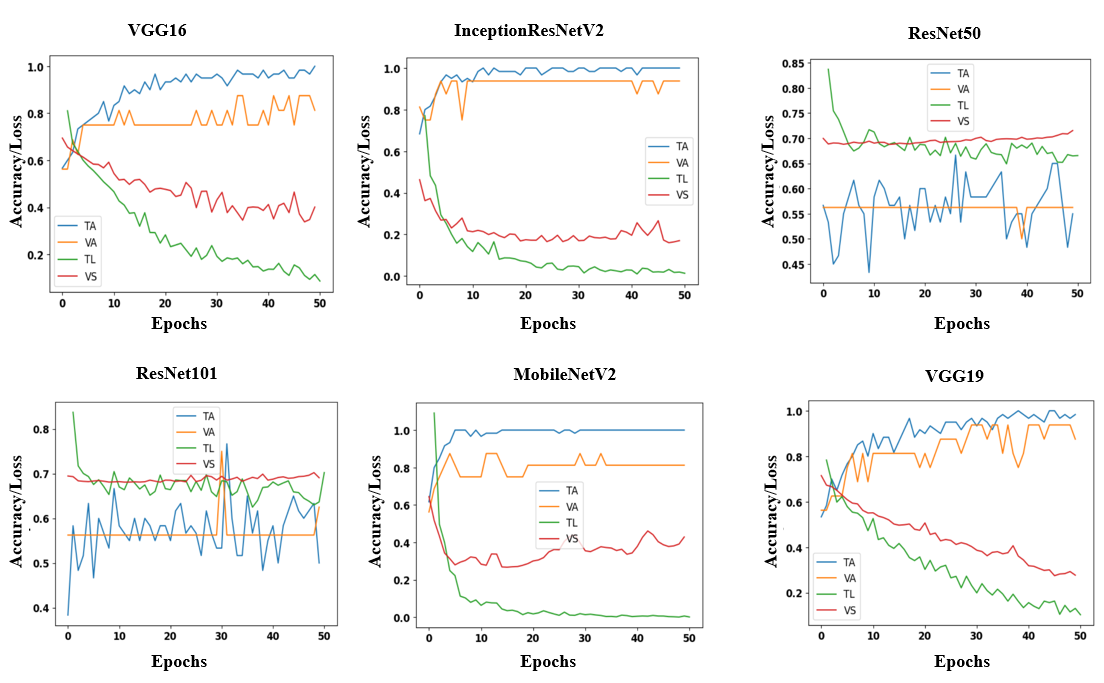}
    \caption{Modified models accuracy and loss during each epoch applied to Study One. TA—train accuracy, VA—validation accuracy, TL—train loss, VS—validation loss.}
    \label{fig:stdtv}
\end{figure*}

Figure~\ref{fig:stdtv} shows train and test performance during each epoch for the CNN models used in this study. The figure shows that the model's performance during each epoch improves for InceptionResNetV2 and VGG19 until 50 epochs. On the other hand, ResNet50 and ResNet101's performance fluctuated during each epoch and overfitted at 50 epochs.

\subsubsection{Explainable AI with Lime}
\begin{figure*}
    \centering
    \includegraphics[width=\textwidth]{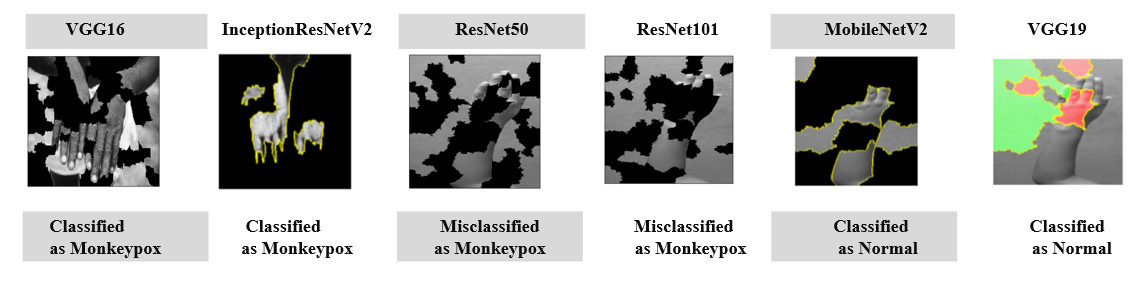}
    \caption{The top four features that facilitated the classification and misclassification of monkeypox patients from infected or non-infected images during Study One.}
    \label{fig:stde}
\end{figure*}

Figure~\ref{fig:stde} explains the model's prediction using LIME. The best performance was observed when LIME model is used with 150 perturbation. The difference between each perturbation was calculated by employing cosine metric with a kernel size of 0.2. A simple linear interpolation is used to provide the explanation of the models. Furthermore, the coefficient was calculated for each superpixel value. Finally, top four features are identified that potentially affect the model's prediction.

\subsection{Study Two}
In Study Two, on the train set (refer to Table~\ref{tab:std2}), InceptionResNetV2 and MobileNetV2 outperformed all other models across all measures, while ResNet50 and ResNet101 demonstrated the worst performance.
\begin{table}[ht]
\caption{Monkeypox prediction performance of all models on training set using various statistical measurements for Study Two.}
    \centering
    \resizebox{\textwidth}{!}{
    \begin{tabular}{ccccccc}\toprule
         \multirow{2}{*}{Model}&	\multicolumn{5}{c}{Performance}\\\cmidrule{3-7}
	&Accuracy&	Precision&	Recall&	F1-Score&	Sensitivity&	Specificity\\\midrule
VGG16&0.97&	0.98&	0.97&	0.98&	0.99&	0.94\\
InceptionResNetV2&
	1&	1&	1&	1&	1&	1\\
ResNet50&0.72&	0.51&	0.72&	0.6&	1&	0\\
ResNet101&	0.72&	0.36&	0.50&	0.42&	1&	0\\
MobileNetV2&	1&	1&	1&	1&	1&	1\\
VGG19&	0.94&	0.94&	0.92&	0.93&	0.97&	0.85\\\bottomrule
 
    \end{tabular}}
    
    \label{tab:std2}
\end{table}

Table~\ref{tab:std2t} shows the test performance of various models applied during study two as a transfer learning approach. The highest performance was observed for MobileNetV2, and the lowest performance was identified for ResNet101 across all measures.
\begin{table}[ht]
\caption{Monkeypox prediction performance of all models on training set using various statistical measurements for Study Two.}
    \centering
    \resizebox{\textwidth}{!}{
    \begin{tabular}{ccccccc}\toprule
         \multirow{2}{*}{Model}&	\multicolumn{5}{c}{Performance}\\\cmidrule{3-7}
	&Accuracy&	Precision&	Recall&	F1-Score&	Sensitivity&	Specificity\\\midrule
VGG16&	0.93&	0.93&	0.90&	0.91&	0.97&	0.82\\
InceptionResNetV2&
	0.98&	0.98&	0.97&	0.98&	0.99&	0.95\\
ResNet50&	0.72&	0.51&	0.72&	0.6&	1&	0\\
ResNet101&	0.72&	0.36&	0.50&	0.42&	1	&0\\
MobileNetV2&0.99&	0.99&	0.99&	0.99&	1&	0.97\\
VGG19&	0.90&	0.89&	0.86&	0.87&	0.95&	0.76\\\bottomrule
  \end{tabular}}
    
    \label{tab:std2t}
\end{table}

\subsubsection{Confusion matrices}
\begin{figure*}[ht]
    \centering
    \includegraphics[width=\textwidth]{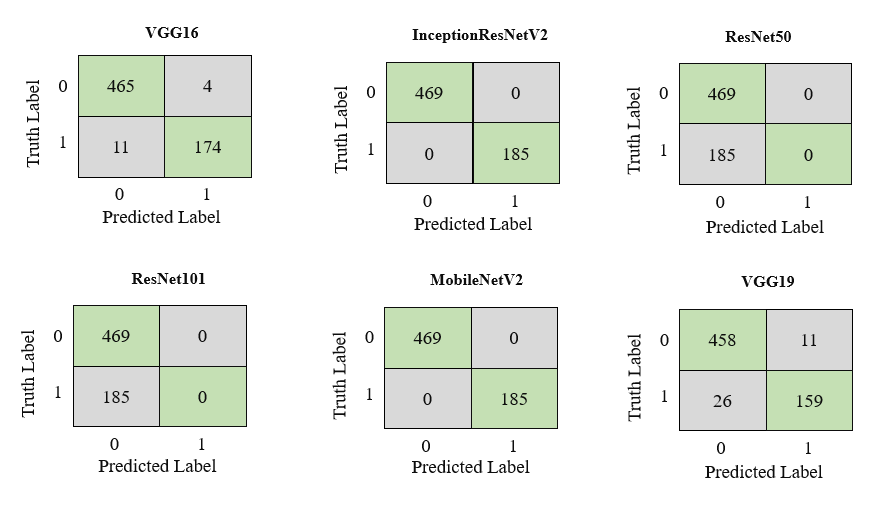}
    \caption{Study Two confusion matrices for the train set.}
    \label{fig:con2}
\end{figure*}

 Figure~\ref{fig:con2} shows the confusion matrices for the train set during Study Two. Both the InceptionResNetV2 and MobileNetV2 accurately categorized all patients, as seen in the Figure~\ref{fig:con2}. In contrast, VGG16, ResNet50, Resnet101 and VGG19 misclassified 15, 185, 185 and 39 patients respectively. As a result, we are able to draw the conclusion that the performance of the ResNet50 and ResNet101 models is the worst out of all of them based on the analysis of the confusion matrices.
 
\begin{figure*}
    \centering
    \includegraphics[width=\textwidth]{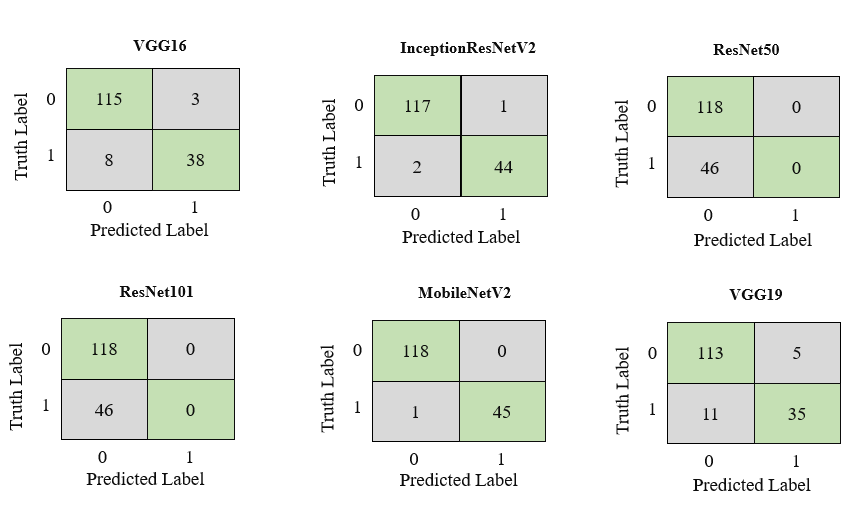}
    \caption{Study Two confusion matrices for the test set.}
    \label{fig:con2t}
\end{figure*}

Figure~\ref{fig:con2t} displays the confusion matrices for Study One on the test set. MobileNetV2 displayed the best performance by misclassifying only one patient, while ResNet50 and ResNet101 indicated the worst performance by misclassifying 46 patients. Considering both train and test analysis, it can be inferred that the best performance of modified MobileNetV2 remains constant for both train and test sets.

\subsubsection{Models’ performance during training}

\begin{figure*}[ht]
    \centering
    \includegraphics[width=\textwidth]{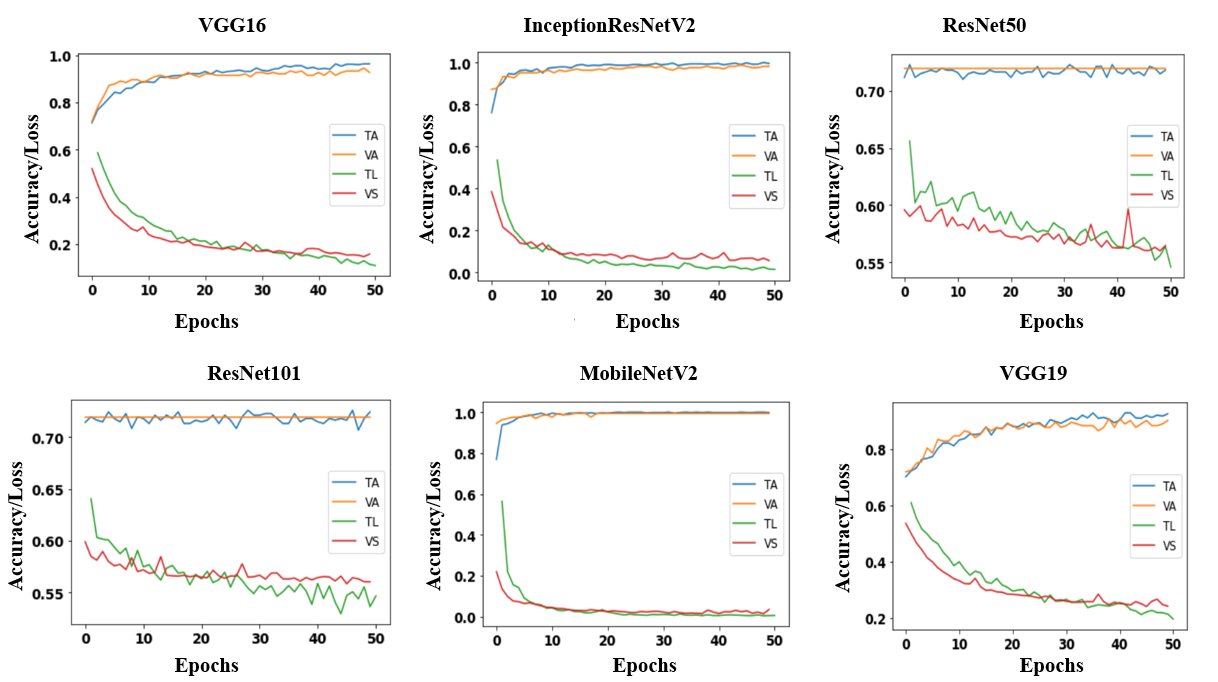}
    \caption{Modified models accuracy and loss during each epoch applied to Study Two. TA—train accuracy, VA—validation accuracy, TL—train loss, VS—validation loss}
    \label{fig:st2}
\end{figure*}

Figure~\ref{fig:st2} plotted train and test results during each epoch for all the CNN models for Study Two. The figure shows that the model's performance during each epoch shows better performance for most of the models except ResNet50 and ResNet101. The accuracy and loss for both the train and test set fluctuated and overfitted for models ResNet50 and ResNet101 after 25 epochs.

\subsubsection{Explainable AI with Lime}
\begin{figure*}[ht]
    \centering
    \includegraphics[width=\textwidth]{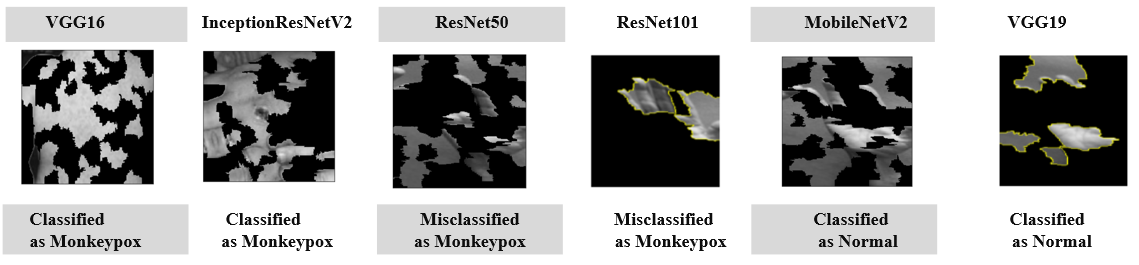}
    \caption{The top four features that facilitated the classification and misclassification of monkeypox patients from infected or non-infected hand images during Study Two.}
    \label{fig:lime2}
\end{figure*}

The model's performance and explanation found in Study one align with Study Two. As shown in Figure~\ref{fig:lime2}, it is clear that ResNet50 and ResNet101 demonstrated the least performance by misclassifying Monkeypox patients.

\section{Discussion}~\label{discussion}


This study proposes and evaluates six distinct modified deep learning-based models in an effort to identify between patients with monkeypox and non-monkeypox symptoms. Our dataset contains limited samples, and we have further evaluated our model's accuracy with 95\% confidence intervals inspired by referenced previous literature~\cite{narin2021automatic,arias2020varicella, sriwong2019dermatological}. Table~\ref{tab:coms1} shows model accuracy for Study One with 95\% confidence intervals for the train and test set. Two approaches have been applied to measure the performance with CI: Wilson score and binomial proportional interval. As stated in Table~\ref{tab:coms1}, InceptionResNetV2 exceeded all other models in terms of accuracy. In contrast, ResNet50 demonstrates significantly lower performance on the train and test set while measuring the accuracy with CI.

\begin{table}[ht]
 \caption{Confidence Interval ($\alpha = 0.05$) of all models used in Study One in terms of accuracy.}
    \centering
    \resizebox{\textwidth}{!}{
    \begin{tabular}{ccccc}\toprule
       Study One&	\multirow{2}{*}{Model}&	\multirow{2}{*}{Train Accuracy}&	Methods\\\cmidrule{4-5}
		&&&	Wilson Score&	Binomial proportion interval\\\midrule
\multirow{6}{*}{Train set}& VGG16&	0.98&	0.851-0.994&	0.914-1.0\\
	&InceptionResNetV2&
	1&	0.93-0.99&	1.00-1.00\\
	&ResNet50&	0.56&	0.44-0.68&	0.44-0.692\\
	&ResNet101&	0.68&	0.55-0.78&	0.56-0.80\\
	&MobileNetV2&	1&	0.93-0.99&	1.0-1.0\\
	&VGG19&	0.98&	0.91-0.99&	0.95-1.0\\\hline
\multirow{6}{*}{Test set}& VGG16&0.81&	0.56-0.93&	0.62-1.0\\
&InceptionResNetV2& 0.87&	0.63-0.96&	0.713-1.0\\

&ResNet50& 0.56&	0.33-0.76&	0.32-0.81\\
&ResNet101& 0.56&	0.33-0.76&	0.32-0.81\\
&MobileNetV2& 0.81&	0.56-0.93&	0.62-1.0\\
&VGG19& 0.93&	0.71-0.98&	0.81-1.0\\\bottomrule
  
\end{tabular}}
\label{tab:coms1}
\end{table}

\begin{table}[ht]
\caption{Confidence Interval ($\alpha = 0.05$) of all models used in Study Two in terms of accuracy.}
    \centering
    \resizebox{\textwidth}{!}{
    \begin{tabular}{ccccc}\toprule
       Study Two&	\multirow{2}{*}{Model}&	\multirow{2}{*}{Train Accuracy}&	Methods\\\cmidrule{4-5}
		&&&	Wilson Score&	Binomial proportion interval\\\midrule
\multirow{6}{*}{Train set}& VGG16&	0.97	&0.95-0.98&	0.95-0.98\\
	&InceptionResNetV2&
	1&	0.99-1.0&	1.0-1.0\\
	&ResNet50&	0.72&	0.68-0.75&	0.68-0.75\\
	&ResNet101&	0.72&	0.68-0.75&	0.68-0.75\\
	&MobileNetV2&	1&	0.99-1.0&	1.0-1.0\\
	&VGG19&	0.94&	0.91-0.95&	0.92-0.95\\\hline
\multirow{6}{*}{Test set}& VGG16&0.93&	0.88-0.96&	0.89-0.97\\
&InceptionResNetV2& 0.98&	0.94-0.99&	0.961-1.0\\

&ResNet50& 0.72&	0.68-0.75&	0.68-0.75\\
&ResNet101& 0.72&	0.68-0.75&	0.68-0.75\\
&MobileNetV2& 0.99&	0.96-0.99&	0.98-1.0\\
&VGG19& 0.90&	0.84-0.93&	0.85-0.94\\\bottomrule
\end{tabular}}
   
\label{tab:comp2}
\end{table}

\begin{figure}[ht]
    \centering
    \includegraphics[width=\textwidth]{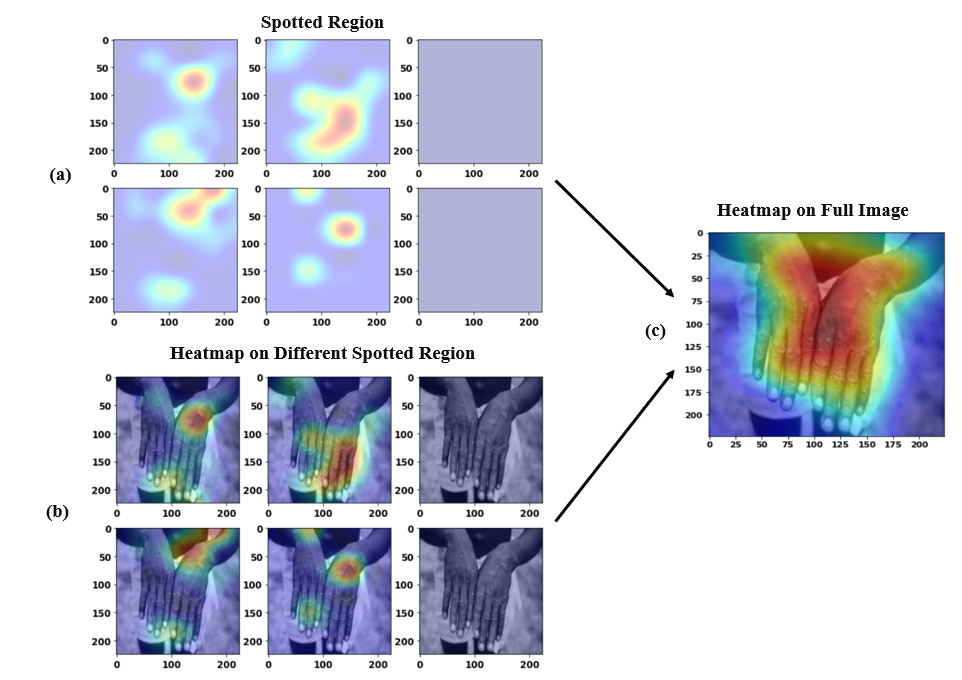}
    \caption{Heatmap of class activation on Monkeypox patient image.}
    \label{fig:hmap}
\end{figure}

Table~\ref{tab:comp2} manifests the accuracy performance of six different models for Study Two. As stated in the table, MobileNetV2 outperformed all other models by showing better performance while ResNet50 and ResNet101 demonstrate the worst performance. Further, we have used Heatmap as a data visualization technique that demonstrates the phenomenon as color in two dimensions for different CNN layers during the prediction. Using heatmap, it is easy to identify the specific region where CNN establishes more focus during the prediction. The idea is to use the weights from the last dense layers and multiply them with the final CNN layer. In this work, we identify the potential Monkeypox infected areas using global average pooling (GAP). Class activation map (CAM) is initially used to collect each output of the convolutional layer and combine it in one shot. Figure~\ref{fig:hmap}(a) represents the spotted region by the ResNet50 models, whereas Figure~\ref{fig:hmap}(b) shows the heatmap on different spotted regions and Figure~\ref{fig:hmap}(c) on  actual images. From the overall analysis across various measures, it can be inferred that MobileNetV2 and InceptionResNetV2 are the best models that can detect Monkeypox patients with perfect accuracy. It is relevant to emphasize that, at the time of experiment, there was no published or preprint literature where the transfer learning approach or deep learning approach was utilized to develop an AI-driven model to detect Monkeypox patients. Consequently, we could not evaluate our model's performance by comparing it with others' work. However, our higher performance results align with many referenced research, which established the potentiality of the transfer learning approach to develop image-based disease diagnosis. Examples include the InceptionResNetV2 model proposed by Narin et al. (2020), which trained with only 100 images to detect COVID-19 patients. Fujisawa et al. (2022) developed a deep learning-based model to detect skin tumors. The model was developed with 1842 patient data and achieved an accuracy of 76.5\%. Our proposed model predictions are explained with LIME and further cross-checked by expert doctors' opinions. Our findings suggest that our modified MobileNetV2 and InceptionResNetV2 demonstrated a significant contribution to identifying the top feature that characterizes the onset of Monkeypox by identifying essential features from the images.

\begin{figure*}[hbt!]
    \centering
\includegraphics[width=.75\textwidth]{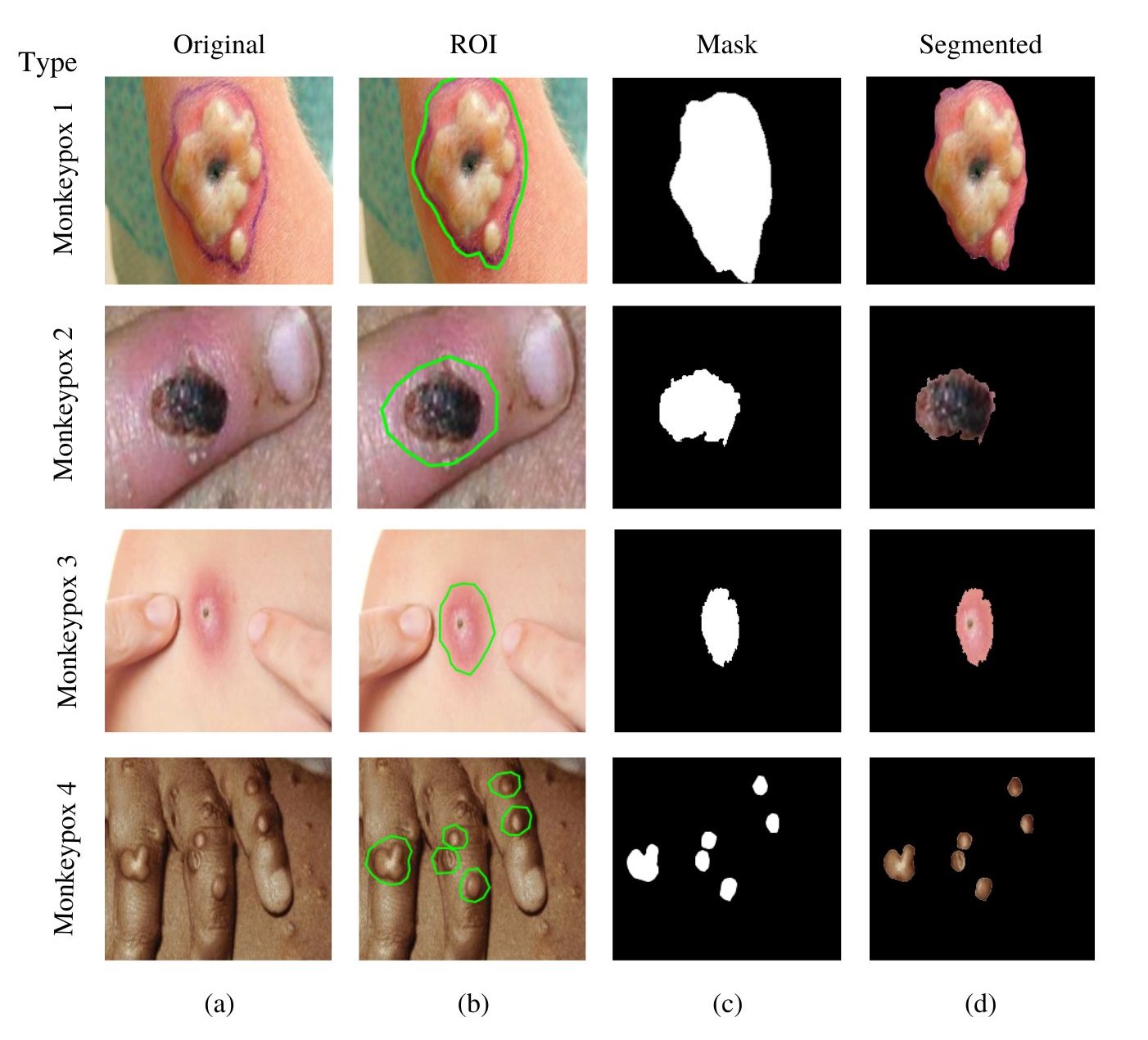}
    \caption{Threshold-based segmentation (a) Original, (b) ROI, (c) Mask, (d) Segmented}
    \label{Segmented}
\end{figure*}

To further aid in comparing the model's final prediction with the visual evidence, we have also employed an image segmentation method to highlight highly infected areas. Figure~\ref{Segmented} depicts the outcomes of threshold-based segmentation applied to images of hands infected with the Monkeypox virus. The region of interest (ROI) has been defined using the pixel values of the infected areas. Then, a mask is generated for use in segmenting and generating the final output.

The major purpose of this research was to create a deep learning-based model capable of detecting patients with Monkeypox symptoms using a dataset of images of Monkeypox patients. Due to a lack of data, the scope of the current literature in this topic remains limited at this time. In the near future, it will be fascinating to examine how our suggested model performs on the multiclass and a large dataset.

\section{Conclusion}\label{conclusion}
This work proposed and tested six alternative modified deep learning models for detecting Monkeypox disease. We discovered that a modified MobileNetV2 and InceptionResNetV2 could identify between Monkeypox and Non-Monkeypox patients with an accuracy ranging from 90\% to 100\%. In addition to that, we have applied LIME, in order to comprehend and verify the predictions made by our model, demonstrating our suggested models' superiority in detecting Monkeypox illnesses. We expect that our study outcomes will provide future academics and practitioners some insight into the use of transfer learning models and explainable AI in constructing secure and trustworthy Monkeypox disease diagnosis models. The steps that are being used during this study will be considered in developing deep learning models using mixed data, evaluating the model's performance on the highly imbalanced dataset, creating fake samples using Generative adversarial networks (GAN), and developing a mobile-based user-friendly application that can be used to diagnose Monkeypox disease.
\bibliographystyle{unsrt}  
\bibliography{main}

\end{document}